\documentclass{article}

\usepackage{arxiv}

\usepackage[utf8]{inputenc} 
\usepackage[T1]{fontenc}    
\usepackage{hyperref}       
\usepackage{url}            
\usepackage{booktabs}       
\usepackage{amsfonts}       
\usepackage{nicefrac}       
\usepackage{microtype}      
\usepackage{lipsum}
\usepackage{graphicx}
\graphicspath{ {./images/} }

\title{MetasurfaceViT: A generic AI model for metasurface inverse design}

\author{
 Jiahao Yan\textsuperscript{*}, Jilong Yi, Churong Ma, Yanjun Bao, Qin Chen, Baojun Li\textsuperscript{*}\\
  Guangdong Provincial Key Laboratory of Nanophotonic Manipulation, \\
  Institute of Nanophotonics,\\
  College of Physics and Optoelectronic Engineering, \\
  Jinan University, \\
  Guangzhou 511443, China\\
  \textsuperscript{*} Corresponding author: \href{mailto:jhyan@jnu.edu.cn}{\texttt{jhyan@jnu.edu.cn}} \& \href{mailto:baojunli@jnu.edu.cn}{\texttt{baojunli@jnu.edu.cn}}
}

\begin{document}
\maketitle
\begin{abstract}
Metasurfaces, sub-wavelength artificial structures, can control light's amplitude, phase, and polarization, enabling applications in efficient imaging, holograms, and sensing.
Recent years, AI has witnessed remarkable progress and spurred scientific discovery.
In metasurface design, optical inverse design has recently emerged as a revolutionary approach.
It uses deep learning to create a nonlinear mapping between optical structures and functions, bypassing time-consuming traditional design and attaining higher accuracy.
Yet, current deep-learning models for optical design face limitations.
They often work only for fixed wavelengths and polarizations, and lack universality as input-output vector size changes may require retraining.
There's also a lack of compatibility across different application scenarios.
This paper introduces MetasurfaceViT, a revolutionary generic AI model.
It leverages a large amount of data using Jones matrices and physics-informed data augmentation.
By pre-training through masking wavelengths and polarization channels, it can reconstruct full-wavelength Jones matrices,
which will be utilized by fine-tuning model to enable inverse design.
Finally, a tandem workflow appended by a forward prediction network is introduced to evaluate performance.
The versatility of MetasurfaceViT with high prediction accuracy will open a new paradigm for optical inverse design.
\end{abstract}

\keywords{Metasurface \and Inverse Design \and Artificial Intelligent \and Vision Transformer}

\section{Introduction}
Optical design plays a crucial role in modern technology, as it allows for the precise manipulation of light, facilitating a wide array of applications such as high-speed communication, high-resolution imaging, and advanced sensing \cite{so2020deep, wiecha2021deep, liu2018generative, peng2024arbitrary, xu2024physics, xi2024deep, zhang2025deep, ma2021deep, ma2022pushing, parkinterfacing}.
In recent years, a revolutionary technique known as optical inverse design has emerged. This approach harnesses the power of deep learning to establish a nonlinear mapping between optical structures and their functional properties \cite{so2020deep, ma2021deep, parkinterfacing, yan2024design}, circumventing the time-consuming traditional design procedures and achieving higher design accuracy \cite{liu2018generative, xu2024physics}. It has found extensive applications, including the design of nanophotonic devices, metasurfaces for beam steering, polarization holograms, and chiral metasurfaces for biosensing \cite{so2020deep, liu2018generative, peng2024arbitrary, xu2024physics, xi2024deep, zhang2025deep}.

Meanwhile, the remarkable advancements and rapid iterations in AI models have reshaped the relationship between AI and science continuously. LLMs, built on the Transformer framework \cite{vaswani2017attention}, have attained human-level intelligence, spurring extensive research on LLM-based autonomous agents for diverse applications \cite{wang2024survey}. The Transformer frameworks have also demonstrated outstanding performances in computer vision \cite{han2022survey}, protein structure prediction \cite{jumper2021highly}, and chemical toxicity assessment \cite{gustavsson2024transformers}. Regardless of whether leveraging LLMs or constructing expertise models, AI has profoundly accelerated scientific discovery \cite{zeni2025generative, wang2023scientific}. As AI models become increasingly sophisticated, the key to enhancing performances has shifted from a model-centric to a data-centric approach \cite{zha2025data}. Scaling laws indicate that large-scale data is essential for optimizing model performance and resource utilization \cite{kaplan2020scaling, zhai2022scaling}.

However, current research on metasurface design using deep learning models has significant limitations. Although considerable progress has been made in enhancing training efficacy and design accuracy, these models typically operate only for fixed wavelengths and polarizations \cite{peng2024transfer, he2024meta, an2022deep}. Transfer-learning and physics-informed training methods have been proposed successfully to achieve wavelength transfer and improve model versatility \cite{zhu2025frequency, xu2024spectral}, yet a change in input-output vector size often necessitates retraining, highlighting the lack of universality in these models.

On the other hand, current research on metasurface inverse design also suffers from a lack of compatibility across different application scenarios. Researchers have made fruitful efforts to enhance model compatibility. For instance, a single physical metric can be utilized to represent multiple problems \cite{sun2024demand, xiong2024deep}; generality can be improved and spectral correlation can be discovered through physical connections \cite{zhang2024harnessing}; the powerful LLM can be employed to fine-tune optical subtasks \cite{kim2025nanophotonic}; or large amounts of data can be prepared to train an "OptGPT" model \cite{ma2024optogpt}. Nevertheless, a versatile and generic design model that can handle diverse applications (encompassing the design of phases and amplitudes for arbitrary wavelengths and polarizations) remains elusive.

In this paper, we introduce MetasurfaceViT, a revolutionary generic AI model for metasurface inverse design. Firstly, to develop a general-purpose large model, we substantially expanded the dataset to 60 million samples. Notably, we innovatively adopted Jones matrices as training data and implemented physics-informed data augmentation, making data acquisition cost-effective. Secondly, inspired by language models and masking pre-training in CV, we pre-trained the model by randomly masking wavelengths and polarization channels. This enables the reconstruction of full-wavelength Jones matrices for various polarization and wavelength combinations, facilitating subsequent designs. Thirdly, based on the pre-trained model for Jones matrix reconstruction, we fine-tuned it to achieve the inverse design from Jones matrix to structural parameters and constructed a forward prediction network to evaluate the performance. Moreover, we devised a workflow for constructing predicted units and evaluating metasurface performance as a whole. Consequently, MetasurfaceViT represents a versatile model/framework in metasurface design, capable of one-shot structure design for arbitrary wavelength, polarization, and application requirements. We verified that the prediction accuracy can exceed 99\% under physically realistic designed amplitudes and phases and reach over 85\% even for ideal ones. Overall, this work is dedicated to constructing a general model for efficient optical inverse design.

\section{Results and Discussion}
\label{sec:headings}

\subsection{Design of wavelength-dependent Jones Matrix \& Masking Strategy}

\begin{figure}
  \centering
  \includegraphics[width=\textwidth,height=\textheight,keepaspectratio]{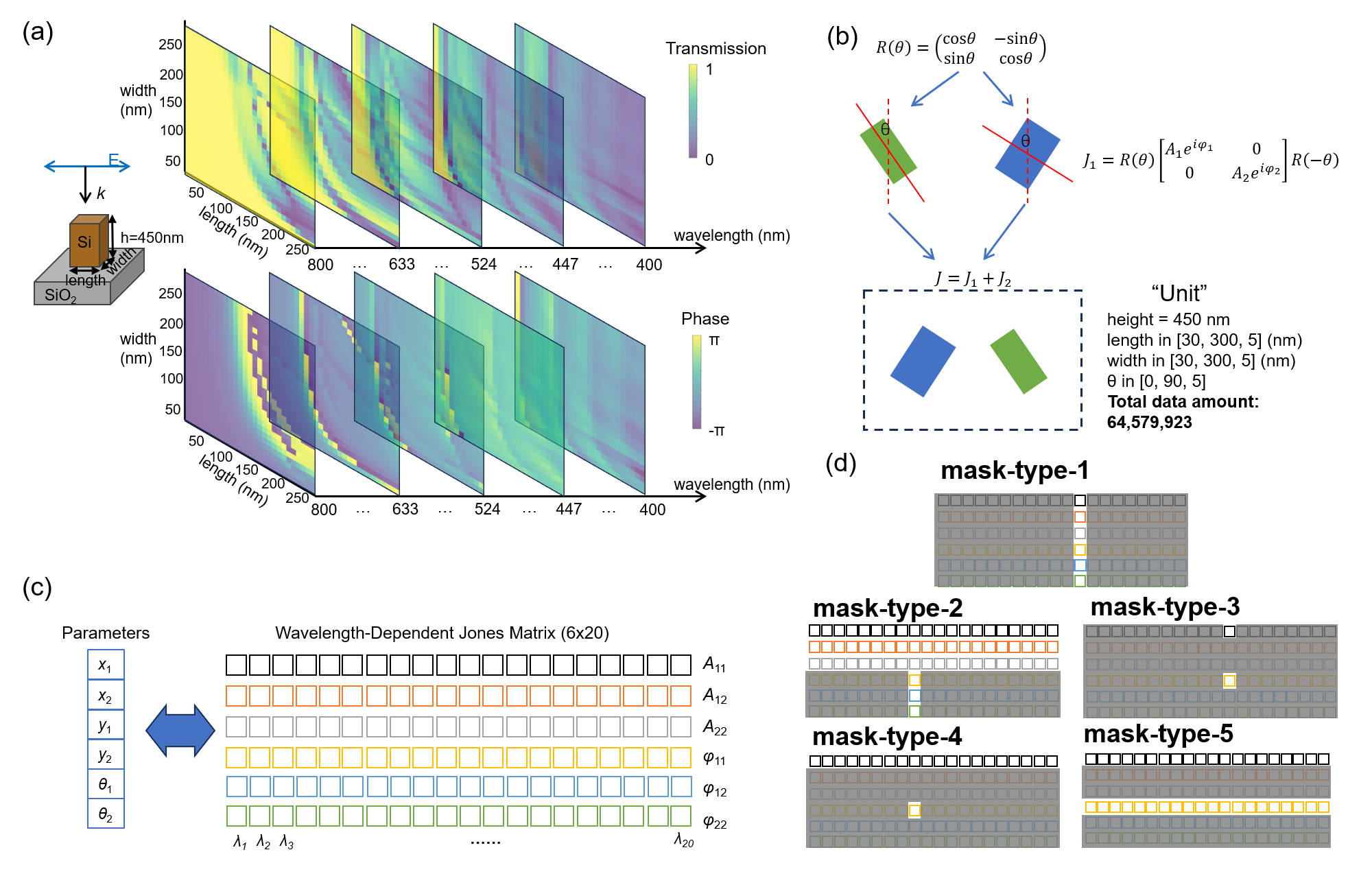}   
  \caption{(a) Schematic illustration of the building block of metasurfaces, which consists of Si nanopillars on a $SiO_2$ substrate. Presented are series of transmission and phase distribution maps, demonstrating how variations in the lengths and widths of the Si nanopillars impact them at different wavelengths.
        (b) Schematic representation of the formation of the “unit cell” in our study through the rotation and pairing of two building blocks. Given the value ranges of widths and lengths, the resultant data size exceeds 60 million.
        (c) The fundamental structure of the dataset. The input data comprises 20×6 Jones Matrices, where 20 represents the number of wavelength points and 6 denotes the six independent components (A11, A12, A22, $\varphi$11, …) at each wavelength. The labels are the six structural parameters of the “unit”.
        (d) Schematics depicting the five strategies employed to mask the Jones matrices during the pre-training phase. Mask-type-1: retain all six components at a randomly selected wavelength; Mask-type-2: preserve amplitudes across all wavelengths while only retaining phases at one randomly chosen wavelength; Mask-type-3: only maintain the amplitude and phase of one component at a single wavelength; Mask-type-4: keep all amplitudes of one component, but only preserve the phase of one component at one wavelength; Mask-type-5: retain all amplitudes and phases of one component.}
  \label{fig:fig1}
\end{figure}

Figure \ref{fig:fig1} illustrates the key components and strategies in our research on AI-enabled optical field manipulation.
In Figure \ref{fig:fig1}a, we depict the building block of metasurfaces, which consists of a Si nanopillar on a $SiO_2$ substrate.
The series of transmission and phase distribution maps are crucial as they reveal the influence of varying lengths and widths of the Si nanopillar on these optical properties at different wavelengths.
This information is fundamental for understanding how the physical structure of the metasurface affects light propagation, which serves as the basis for our subsequent data-driven approaches.
Figure \ref{fig:fig1}b shows the concept of constructing a \"unit cell\" by rotating and pairing up two building blocks.
The large data size, exceeding 60 million, is a result of the wide value ranges of widths, lengths, and angles considered.
This large dataset provides a rich source of information for training our AI models, enabling us to explore a vast parameter space and potentially discover novel optical regulation phenomena.
The dataset structure is presented in Figure \ref{fig:fig1}c. We use 20x6 Jones Matrices as input data, where the 20 represents the number of wavelength points in the visible light region, and the 6 corresponds to the three amplitude components and three phase components of the Jones matrix at each wavelength.
The six structural parameters of the \"unit\" are used as labels. This mapping between the input data and labels is the cornerstone of our training process, as it allows our AI models to learn the relationship between the physical structure and the optical response.
Figure \ref{fig:fig1}d details the five masking strategies employed during the pre-training phase.
These strategies are designed to enhance the generalization ability of our models.
For example, Mask-type-1 keeps all six components at a random wavelength, which helps the model to learn the characteristics of a single wavelength while ignoring others, potentially improving its ability to handle different wavelengths separately.
Mask-type-2 focuses on amplitudes across all wavelengths and phases at a single random wavelength, emphasizing the importance of amplitude information while still considering phase information at one point.
Mask-type-3 only retains the amplitude and phase of one component at one wavelength, which can force the model to focus on specific aspects of the optical response.
Mask-type-4 and Mask-type-5 further refine the focus on components and wavelengths, respectively.
Overall, these figures and strategies are integral to our goal of achieving universal optical field regulation.
By using Jones matrix data as training data and applying various masking strategies, we aim to develop AI models that can perform inverse design for any given wavelength and polarization requirements, which is a significant step forward in the field of AI + Optics.


\subsection{Workflow of Pretrain, Design, and Reconstruction}
\begin{figure}
  \centering
  \includegraphics[width=\textwidth,height=\textheight,keepaspectratio]{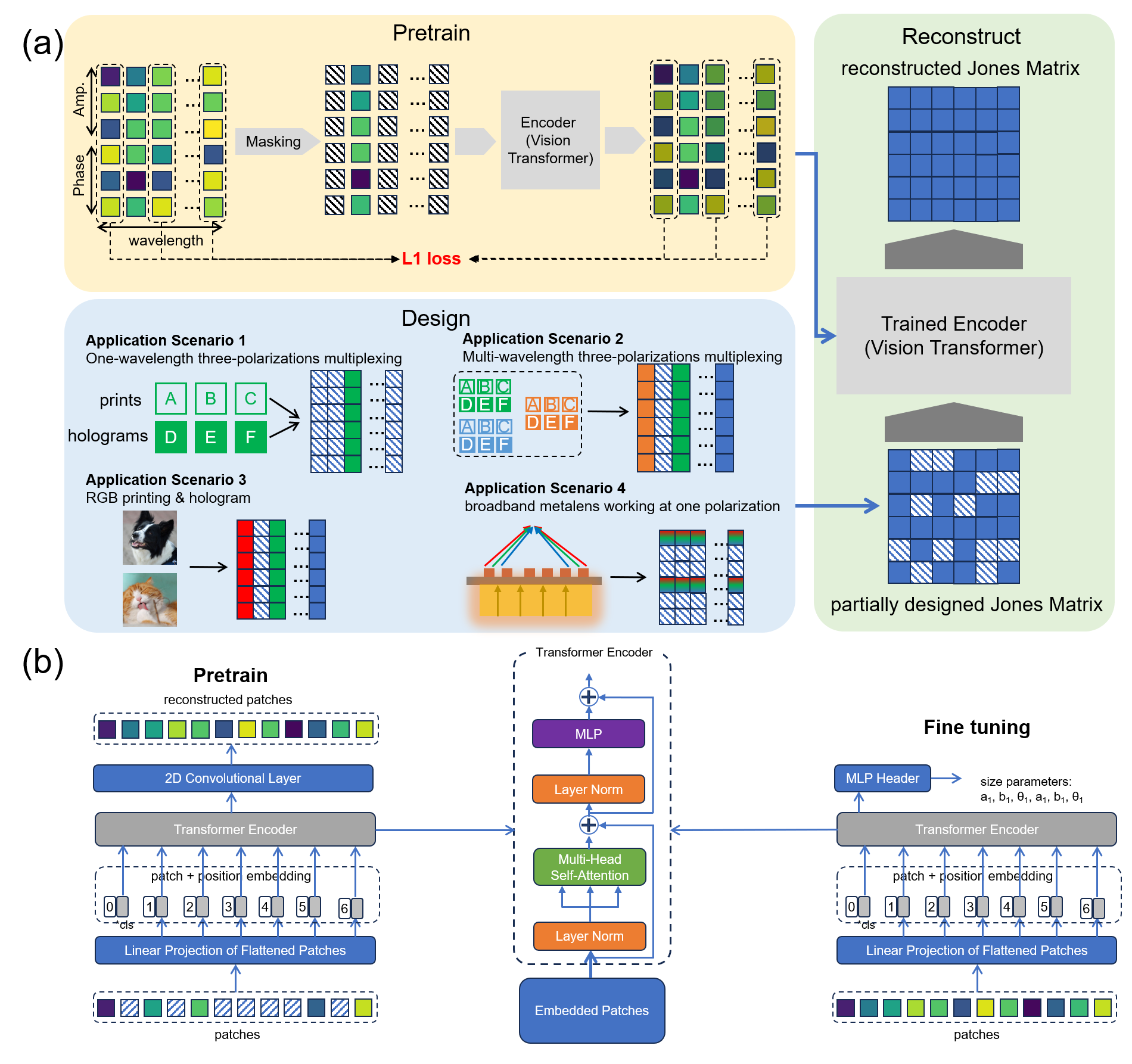}   
  \caption{(a) Workflow of the pretraining, design, and reconstruction phases. Pretraining: Jones Matrices from the training set are masked using different masking strategies and then input into the Vision Transformer (ViT) network. The network is trained by minimizing the L1 loss between the generated values and the actual values of the masked elements. Design: Generate the corresponding Jones Matrix based on diverse application scenarios and populate the 20×6 full-scale Jones Matrices. As long as your application scenario can be manifested and represented by at least one “pixel” within the 20×6 Jones Matrices, it can be addressed. Typical application scenarios include three-polarizations multiplexing at single or multiple wavelengths, RGB holograms, routing or printing, and broadband metalenses. Reconstruction: Utilize the incomplete Jones Matrices from the design phase and feed them into the pre-trained ViT network. Eventually, complete Jones Matrices with filled blank elements can be obtained.
      (b) Architecture of the ViT network during the pretraining and fine-tuning phases.}
  \label{fig:fig2}
\end{figure}

Figure \ref{fig:fig2} illustrates the workflow and architecture of a method that combines AI and optics, specifically focusing on the use of a Vision Transformer (ViT) in the context of Jones Matrix manipulation.
In the pretrain phase, as depicted in the upper-left part of Figure \ref{fig:fig2}a, Jones Matrices from the training set are masked using various masking strategies.
These masked matrices are then fed into the ViT network. The goal of this training process is to minimize the L1 loss between the generated values and the actual values of the masked elements.
This optimization step through gradient descent is crucial for training the network to accurately predict the masked parts of the Jones Matrices.

The design phase, shown in the middle section of Figure \ref{fig:fig2}a, maps user requirements for metasurfaces to Jones Matrices.
Different application scenarios \cite{bao2022observation, bao2025single, bao2021toward}, such as one-wavelength three-polarizations multiplexing, multi-wavelength three-polarizations multiplexing, RGB printing \& holograms, and broadband metalens working at one polarization, are considered.
The corresponding Jones Matrices are generated and partially filled into a 20x6 full-scale Jones Matrix.
This process allows for the representation of diverse optical applications within a unified matrix framework.

The reconstruction phase, in the upper-right part of Figure \ref{fig:fig2}a, takes the partially filled Jones Matrices from the design phase and passes them through the pre-trained ViT network.
The network then reconstructs the blank elements, resulting in a complete Jones Matrix.
This complete matrix is essential for subsequent processing in the finetune workflow.

Figure \ref{fig:fig2}b presents the architecture of the ViT network during the pretrain and finetuning phases.
In the pretrain phase, the network processes reconstructed patches through a series of operations including a 2D convolutional layer, a linear projection of flattened patches, and the ViT encoder itself.
The ViT encoder consists of a multi-head self-attention mechanism, layer normalization, and a multi-layer perceptron (MLP).
During the finetuning phase, the network output is adjusted to generate size parameters.
The key difference between the pretrain and finetune phases lies in the output of the network.
While the pretrain phase focuses on reconstructing the Jones Matrix, the finetune phase aims to optimize specific parameters related to the application.

This combination of pretrain, design, reconstruction, and finetuning processes using a ViT network in the context of Jones Matrices provides a powerful framework for solving various optical problems.
By leveraging the capabilities of AI, it enables efficient manipulation and prediction of optical properties encoded in Jones Matrices, which can lead to advancements in optical design, metasurface engineering, and related fields.

\subsection{Workflow of Finetuning, Prediction, and Evaluation}

\begin{figure}
  \centering
  \includegraphics[width=\textwidth,height=\textheight,keepaspectratio]{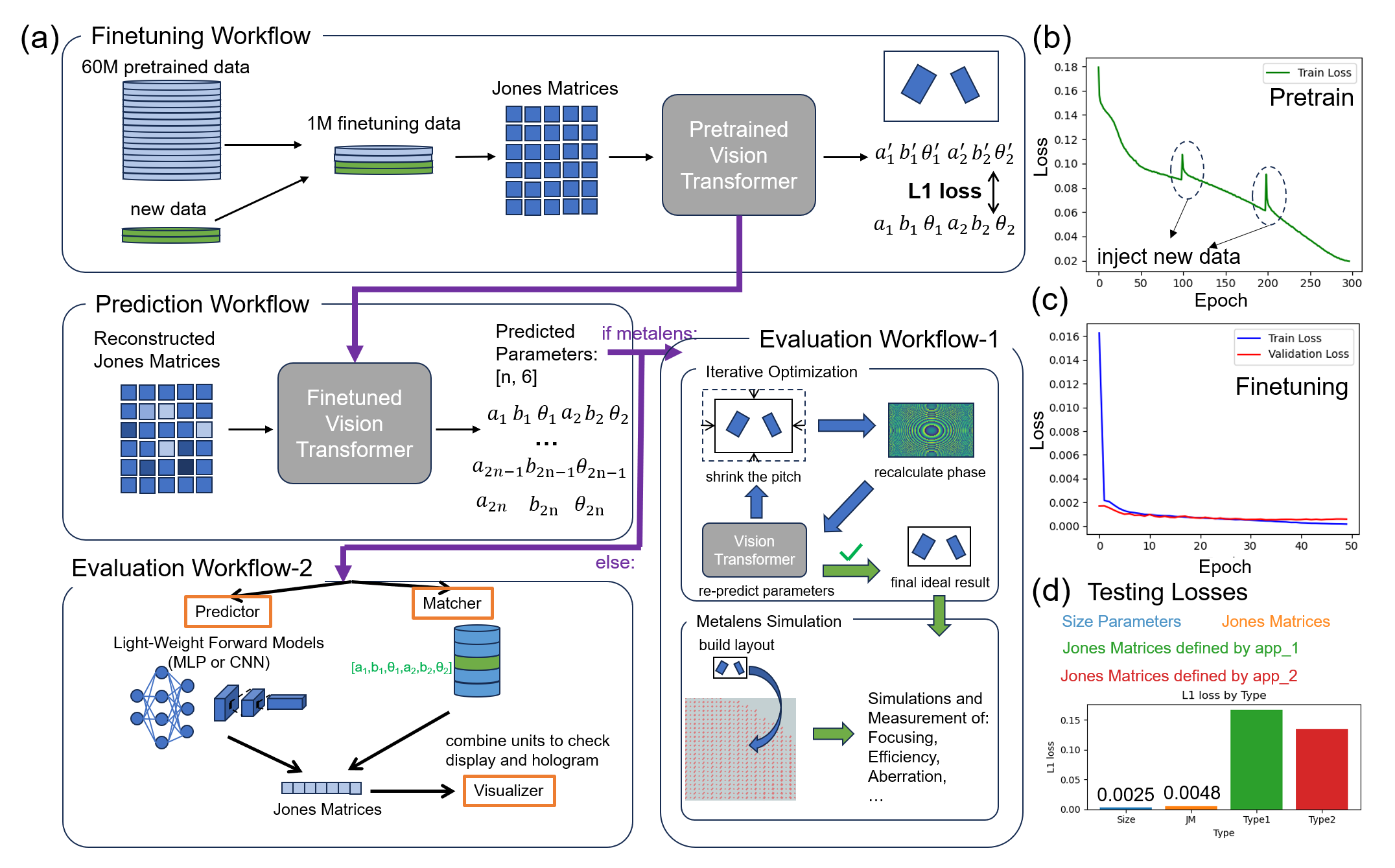}   
  \caption{
(a) Workflows of the fine-tuning, prediction, and evaluation phases. When the design type is a metalens, Evaluation Workflow-1 is initiated. This workflow encompasses iterative size and pitch generation, followed by FDTD-based metalens simulation. In other scenarios, Evaluation Workflow-2 is activated, which consists of a predictor, a matcher, and a visualizer.
(b) The epoch-dependent L1 loss during the pretraining phase. Notably, the two peaks observed at epoch = 100 and 200 result from the data splitting and distributed injection techniques implemented to circumvent memory overhead.
(c) The epoch-dependent L1 loss during the fine-tuning phase. Given that the fine-tuning task is considerably less complex than pretraining, the model can attain a very low loss after only 50 epochs.
(d) Evaluation of Jones Matrices (JM) from the test set and designed Jones Matrices. “Size”: Represents the loss of the inverse design process (JM -> size parameters). “JM”: Denotes the loss of the tandem workflow (JM -> size parameters -> JM). “Type1”: Corresponds to the L1 loss between the designed JM (not for metalens) and the predicted JM through the tandem workflow. “Type2”: Signifies the L1 loss between the designed JM (specified for metalens) and the predicted JM through the tandem workflow.
}
  \label{fig:fig3}
\end{figure}

Figure \ref{fig:fig3} illustrates the workflows of finetuning, prediction, and evaluation phases, along with associated loss metrics during different training and testing stages.
The finetuning workflow (Figure \ref{fig:fig3}a) starts with a combination of 60M pretrained data and new data, which are randomly sampled to form a 1M finetuning dataset.
The pretrained ViT serves as the starting point, and the finetuning process adjusts the network weights.
Unlike the pretraining phase, the loss function in finetuning is the L1 loss between the target and generated structural parameters.
This approach allows the model to adapt more precisely to the specific requirements of the new data related to Jones Matrices.

In the prediction workflow, reconstructed Jones Matrices, generated using the pretrained network, are fed into the finetuned ViT. The output is a set of nx6 structural parameters, where n represents the number of structures.
The remarkable speed of this generation, such as for a 256x256 metasurface with n = 65536, which can be completed in seconds, showcases the efficiency of the model.
The evaluation workflow is crucial for assessing the rationality of the generated structural parameters.
It can be seen as a forward-prediction process, simulating optical effects before the experimental fabrication of metasurfaces.
Evaluation Workflow-1 is specifically designed for metalens, which involves iterative optimization due to the sensitivity of metalens phase design to the absolute position of units and the need to reduce the pitch between units for efficiency.
By iteratively shrinking the pitch and recalculating the phase distribution, the process aims to achieve an optimal design while avoiding excessive fabrication difficulty (with a minimum gap of 100nm).
Evaluation Workflow-2, on the other hand, is for conventional metasurfaces, using a predictor (a lightweight trained network) and a matcher (searching in pre-trained data) to map structural parameters to Jones Matrices, followed by visualization steps.

In Figure \ref{fig:fig3}b, the epoch-dependent L1 loss during the pretraining phase is shown.
The two peaks at epoch = 100 and 200 are due to data splitting and distributed injection to manage memory overhead.
This indicates the challenges in handling large-scale data during pretraining.
Figure \ref{fig:fig3}c depicts the epoch-dependent L1 loss during the finetuning phase. The model can achieve a very low loss after 50 epochs, highlighting that the finetuning task is relatively easier than pretraining, likely because the model already has a good initial set of weights from pretraining.

Figure \ref{fig:fig3}d evaluates the performance on Jones Matrices from the test set and designed Jones Matrices. The low L1 loss of 0.0025 for the inverse design (Jones Matrices -> size parameters) and 0.0048 for the tandem workflow (Jones Matrices -> size parameters -> Jones Matrices) on the test set demonstrates the effectiveness of the model.
However, the higher losses for designed Jones Matrices (around 0.15 for non-metalens and 0.13 for metalens) suggest that the ideal designed matrices may not be fully realizable in practice, as they represent an overly simplified or theoretical scenario.
The iterative and efficient nature of the workflows, along with the analysis of loss metrics, contribute to the understanding of how the model can be optimized for practical applications in metasurface design, while also highlighting the challenges in bridging the gap between theoretical designs and real-world implementation.

\subsection{Example Application 1: Multiplexing of printings and holograms}

\begin{figure}
  \centering
  \includegraphics[width=\textwidth,height=\textheight,keepaspectratio]{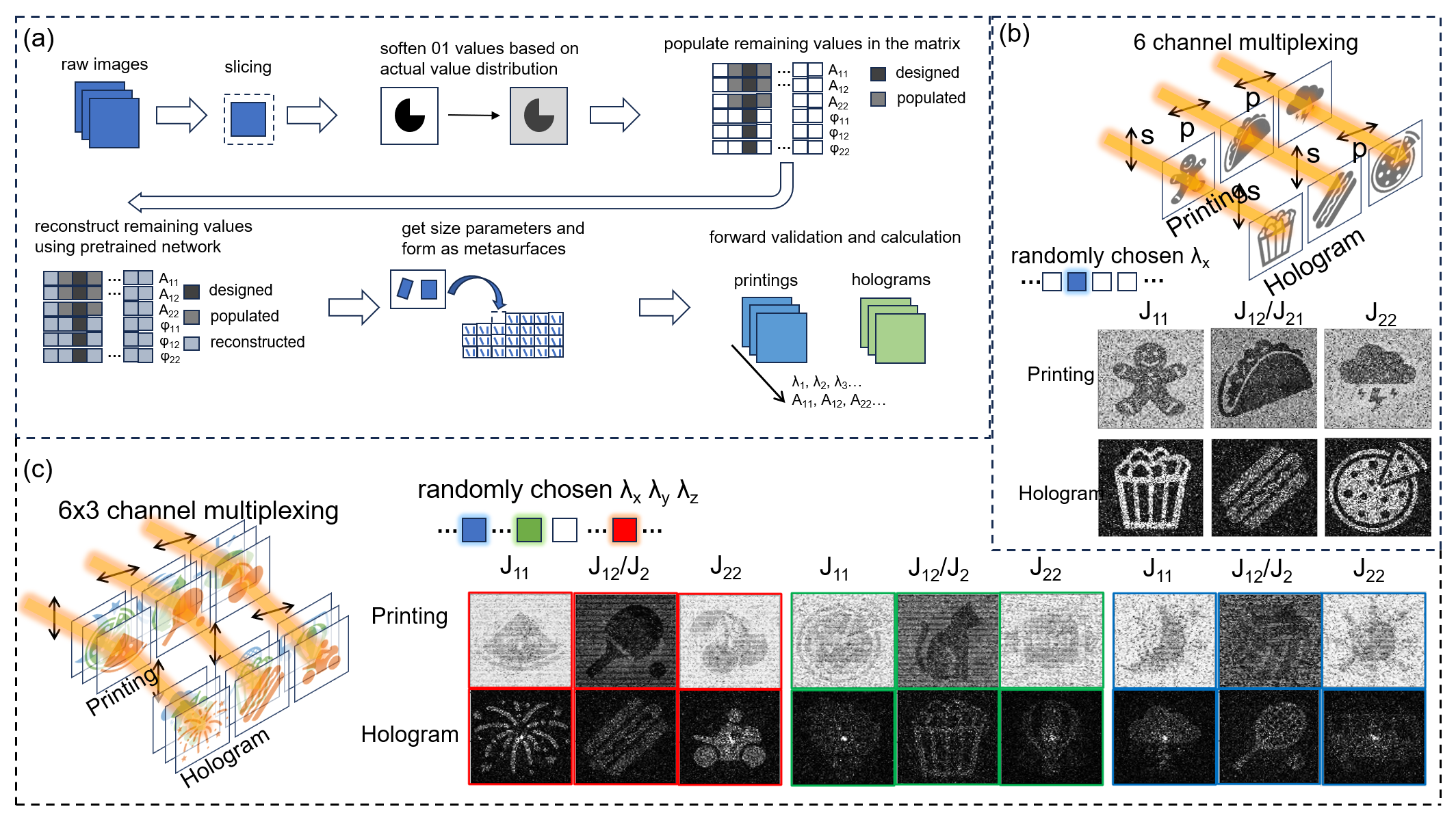}   
  \caption{(a) Schematic illustration of the process for designing and evaluating multiplexing metasurfaces.
(b) Demonstration of 6-channel printing and hologram multiplexing achieved at a single wavelength.
(c) Illustration of 18-channel printing and hologram multiplexing accomplished at three wavelengths.}
  \label{fig:fig4}
\end{figure}

Figure \ref{fig:fig4} illustrates the workflow for designing metasurfaces for multiplexing printings and holograms, as well as the example results of single- and multi-wavelength multiplexing.
In Figure \ref{fig:fig4}a, the process begins with raw images, which are sliced to obtain target images of appropriate sizes.
The binary values in the images are then softened based on the actual value distribution.
For example, the values of 0 and 1 are adjusted according to the amplitude distribution at the corresponding wavelength to ensure that the designed Jones matrix conforms to the actual values.
After designing the amplitude and phase for a certain wavelength, the values of other relevant wavelengths need to be populated.
For instance, when designing the red part of the RGB channels, the amplitude of the red part is kept normal, while the blue and green parts are assigned minimum values to minimize the interference of the red unit on other channels.
The remaining part of the Jones matrix is reconstructed using the pretrained network, and then the size information of the unit is obtained through a finetune network.
This operation is performed on each unit of the metasurface, and the size information is used to generate a predicted Jones matrix through a forward network.
The amplitude and phase are calculated according to the corresponding components to evaluate the effects of printings and holograms.

Figure \ref{fig:fig4}b shows the six-channel printing and hologram multiplexing at one wavelength.
By encoding information in A11, A12, A22, $\varphi$11, $\varphi$12, $\varphi$22, six-channel information multiplexing can be achieved.
As shown in the schematic, by changing the polarization directions of the polarizer and analyzer, multiple channels of information can be read out.
The resulting printings and holograms demonstrate good performance, highlighting the flexibility of the model in reverse-designing at any selected wavelength.

Figure \ref{fig:fig4}c presents the 18-channel printing and hologram multiplexing at three wavelengths.
Similar to the single-channel case, six-component information multiplexing is achieved at each of the three randomly selected wavelengths.
Although the 6×3 channel multiplexing is basically achieved, some holograms suffer from high noise levels, making it difficult to clearly identify the images.
This is due to the idealized design that cannot be fully reproduced, such as the overlapping of the designed hologram images in the three channels, which causes crosstalk.
Overall, this application showcases the versatility and functionality of the model, allowing for inverse design at any wavelength and polarization.
The previous loss information has proven that accurate design can be achieved as long as the designed Jones matrix is close to the physical reality.
Future improvements can be made by considering more optical dependencies in the design phase to obtain better results.

\subsection{Example Application 2: Broadband achormatic metalens}

\begin{figure}
  \centering
  \includegraphics[width=\textwidth,height=\textheight,keepaspectratio]{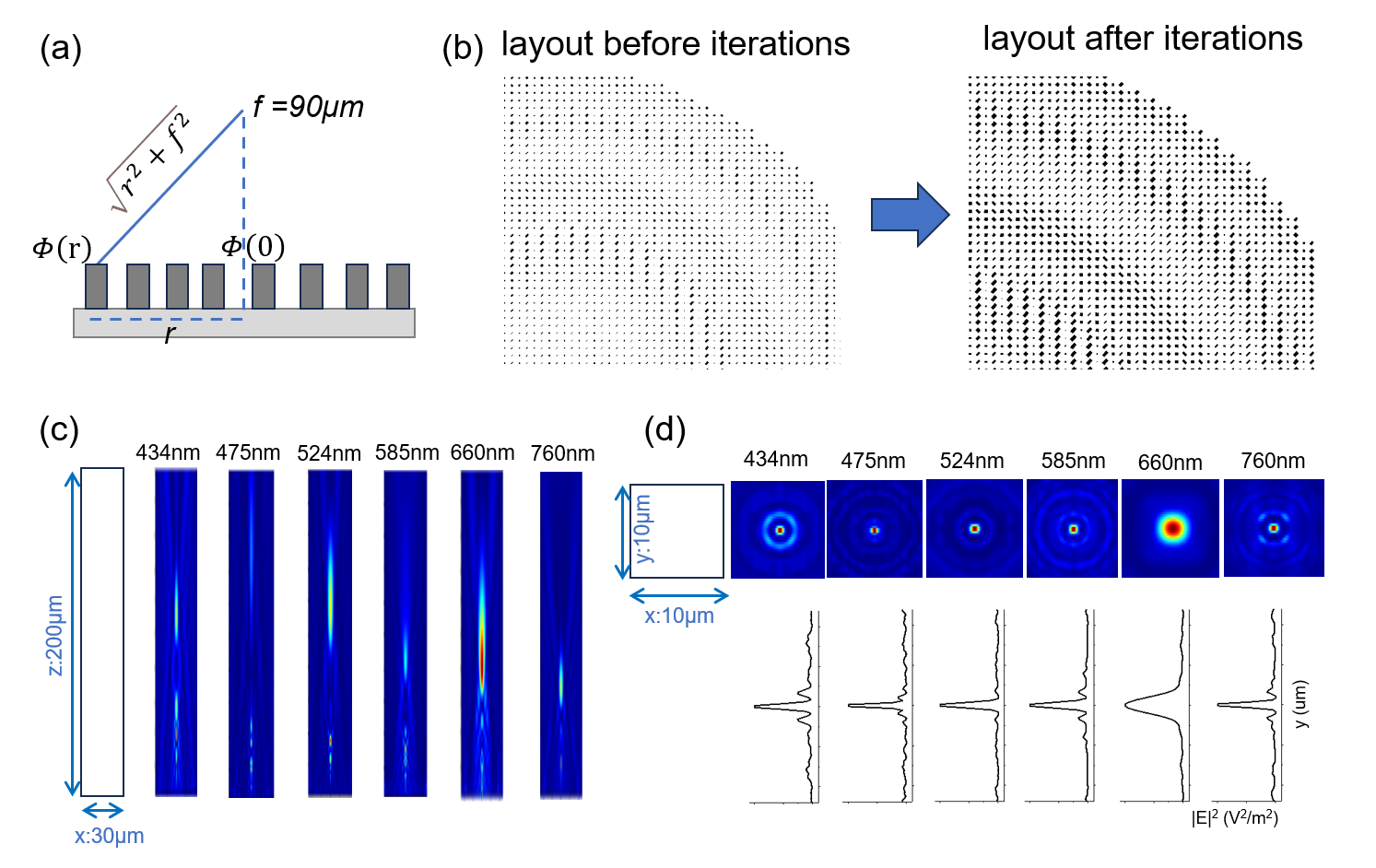}   
  \caption{(a) Schematic representation of the calculation methodology for phase distribution.
(b) Layouts of the metalens before and after iterative optimization, utilized to enhance the inter-unit distance.
(c) Normalized electric field profiles within the x-z plane, indicative of the focusing length corresponding to each wavelength.
(d) Normalized electric field profiles in the x-y plane and line-scan electric field intensities along the y-axis, illustrating the focusing performance at each wavelength.}
  \label{fig:fig5}
\end{figure}

Figure \ref{fig:fig5} is used to evaluate the broadband achromatic metalens designed through our proposed model.
In Figure \ref{fig:fig5}a, a schematic illustration shows the method of calculating the phase distribution of each unit in the metalens.
The phase calculation is based on the distance of each unit from the origin ($r$) and the designed focal length ($f = 90\mu m$).
This approach is fundamental for achieving the desired optical performance of the metalens.
Figure \ref{fig:fig5}b presents the layouts of the metalens before and after iterative optimization.
The initial layout, before iterations, shows a relatively sparse arrangement of metalens units, which is less favorable for interaction with the light field.
In contrast, the layout after iterations is more densely packed while still achieving the preset phase distribution.
This demonstrates the significance of the iterative algorithm in optimizing the distance between units, enhancing the efficiency of light manipulation.

Figure \ref{fig:fig5}c displays the normalized electric field profiles at the x-z plane, which indicate the focusing performance of the metalens across six representative wavelengths in the visible light spectrum (434nm, 475nm, 524nm, 585nm, 660nm, 760nm).
The 3D simulation, though time-consuming, accurately represents the light field.
The focusing length is approximately 90$\mu$m as designed, but a chromatic effect is observed, where the focusing length decreases with the increase of wavelength. This chromatic aberration is an area for improvement.
Figure \ref{fig:fig5}d shows the normalized electric field profiles at the x-y plane and the line-scan electric field intensities at the y-axis.
It reveals that the metalens has good focusing performance at all six wavelengths, with the full-width at half-maximum (FWHM) generally within 1$\mu$m, meeting the basic practical requirements.
Regarding the less-than-ideal achromatic performance, it is mainly attributed to the strong amplitude constraints during the design phase.
When all wavelength amplitude components are set to the maximum value and the phase distribution is applied to all phase components, it may sacrifice some phase constraints.
This can lead to polarization in the actual predicted Jones matrix.
To address this issue, further optimization of the design workflow is necessary, which will be a key area of future research.
Overall, the results provide valuable insights into the performance and limitations of the broadband achromatic metalens, guiding future improvements in its design and optimization.

\section{Conclusion}
\label{sec:others}
In conclusion, MetasurfaceViT emerges as a game-changing force in metasurface inverse design.
By enlarging the dataset to 60 million samples and introducing Jones matrices and physics-informed augmentation, we found a cost-effective way to build large AI model in optics.
The novel pre-training approach, involving random masking of wavelengths and polarization channels, empowered the model to handle diverse combinations adeptly.
Fine-tuning it for the transition from Jones matrix to structural parameters and building a forward prediction network further enhanced its capabilities.
Our verification confirmed remarkable prediction accuracies, surpassing 99\% under realistic conditions and reaching over 85\% for ideal ones.
MetasurfaceViT thus paves the way for efficient, one-shot metasurface structure design, fulfilling arbitrary wavelength, polarization, and application demands, heralding a new era in this domain.

\section{Methods}

\textbf{1. Code Availability:}

The code used in this research is publicly available on GitHub at \url{https://github.com/JYJiahaoYan/MetasurfaceVIT}.
This repository encompasses all the code related to AI and optical calculations employed in the study.
It is structured into several key components: data generation, masked pre-training, metasurface design, Jones matrix reconstruction, fine-tuning of pre-trained models, parameter prediction, and metasurface forward prediction.

\textbf{2. Deep learning model:}

The training process was carried out on a workstation equipped with four NVIDIA 4090 GPUs, leveraging the PyTorch distributed training module for enhanced efficiency.
Each GPU was assigned a batch size of 128, and consequently, with four GPUs operating in parallel, the overall batch size amounted to 512.
When dealing with data on the scale of approximately 60 million samples, one epoch of training took roughly 1 hour.
For a pre-training phase involving 300 epochs, the estimated time consumption was close to 12.5 days, which fell within an acceptable range.
During the fine-tuning stage, due to the substantially reduced data volume and the requirement of only 50 epochs, the entire fine-tuning process took approximately 5 hours.
Notably, the fine-tuning operation could also be performed on an ordinary computer, which significantly broadened the practical applicability of the model.
This study was based on the ViT architecture, incorporated the standard Multi-Head Self-Attention mechanism, with the number of heads set to 12 to capture feature information from diverse dimensions.
In the model training process, the AdamW optimizer was employed, with the initial learning rate set to 1e-4 and the weight decay set to 0.05 to effectively prevent overfitting.
Meanwhile, a Cosine Annealing Learning Rate Scheduler was utilized to dynamically adjust the learning rate and facilitate model convergence.
One difference from conventional ViT is we don't divide input into patches because 20x6 Jones Matrices is simple enough for network.
Therefore, every unit is liearnly projected into the embedding space.

\textbf{3. Optical-related calculation and simulation:}

Firstly, FDTD simulations were employed to gather the phase and transmission data of single silicon nanopillars as their sizes varied.
Then, the Jones matrix for a single structure was built using the amplitude and phase under two orthogonal polarizations, and through matrix operations, the Jones matrices for two rotated nanopillars were obtained.
This construction was performed across all wavelength points, ultimately yielding a 20x6 2D matrix as training data (20 for wavelength points and 6 for 3 amplitude and 3 phase components).
In the design workflow, target images for printings were directly converted to numpy arrays via PIL methods.
For phase design, the Gerchberg-Saxton algorithm was used for iterative optimization via multiple Fourier transforms and inverse Fourier transforms to derive the phase distribution on the metasurface exit plane from target hologram images.
For evaluation workflow-1 (metalens assessment), a 3D metasurface was constructed with FDTD simulations.
Array truncation was pre-applied to ensure a 30 µm metalens diameter, and only one quadrant was built by leveraging symmetry to save computation.
The far-field projection and focusing effects were evaluated via direct simulation for higher accuracy despite longer runtime.
For evaluation workflow-2 (multiplexed metasurface assessment), when multiple wavelength channels were involved, numpy arrays were split and recombined to extract wavelength-specific images from the Jones matrix.
Holograms were generated using the fast Fourier transform.

\section{Acknowledgements}

The work was supported by the Guangdong Basic and Applied Basic Research Foundation (No. 2023B1515020046, 2022B1515020019) and
National Natural Science Foundation of China (62475100).

\bibliographystyle{unsrt}  


\end{document}